\documentclass{jltp}

\usepackage{graphicx} % uncomment this line to include the graphicx package
\title{Aluminum Single Electron Transistors with Islands Isolated from a Substrate}
\author{V.A. Krupenin$^1$, D.E. Presnov$^{1,3}$, A.B. Zorin$^{2,3}$ and J.
Niemeyer$^2$}
\address{$^1$Laboratory of Cryoelectronics, Moscow State University, 119899 Moscow,
Russia\\ $^2$PTB, Bundesallee 100, D-38116 Braunschweig, Germany\\
$^3$Nuclear Physics Institute, Moscow State University, 119899
Moscow, Russia}

\runninghead{\bf V.A. Krupenin {\boldmath $et\ al.$}}{Al SET
Transistors with Islands Isolated from a Substrate}
\begin{document}

\maketitle

\begin{abstract}
The low-frequency noise figures of single-electron transistors
(electrometers) of traditional planar and new stacked geometry
were compared.  We observed a correlation between the charge noise
and the contact area of the transistor island with a dielectric
substrate in the set of Al transistors located on the same chip
and having almost similar electric parameters. We have found that
the smaller the contact area the lower the noise level of the
transistor. The lowest noise value (${\delta}Q_x=(8\pm 2)\times
10^{-6}e/\sqrt{\mbox{Hz}}$ at 10Hz) has been measured in a stacked
transistor with an island which was completely isolated from a
substrate. Our measurements have unambiguously indicated that the
dominant source of the background charge fluctuations is
associated with a dielectric substrate.

PACS numbers: 73.40.Gk, 73.40.Rw
\end{abstract}

%Include this space if you do not use sections in your document.
%\vspace{0.3in}

\section{INTRODUCTION}

Numerous experiments with the Single Electron Tunneling (SET)
devices have shown that the offset charge noise caused by
fluctuations of the polarization charge on small conductive
islands located between the tunnel junctions presents a serious
problem\cite{Zimm,Verb,Zor}. These background charge fluctuations
dominate over the intrinsic shot noise of SET devices\cite{kor1}
at low frequencies ($f<0.1- 1$kHz). They considerably impair
the performance of SET electrometers, pumps\cite{Kel},
traps\cite{vkrup1}, etc. and do not allow the concept
of digital SET devices to be developed \cite{Likh}.

For example, in a metallic SET structure the transistor island is
in contact with the dielectric surrounding, namely the substrate
from a bottom, the tunnel barrier layers and, as in the case of
Al, the native oxide on the open surface of the structure. This
surrounding apparently contains many trapping centers capable of
producing random low-frequency variations of the polarization
charge on the transistor island.

The idea of our study has been to effectively diminish the substrate
component of noise in a SET transistor. We solved this problem by
using the non-traditional (stacked) design of the transistor, in
which the contact area of the island to the substrate was
minimized\cite{vkrup2}. A small metallic island of the SET
transistor was placed onto the oxidized base electrode(s), which
efficiently screened the electric field of charge impurities
located in the substrate.
\section{FABRICATION OF THE SAMPLES}
The Al transistors (see Fig.~1) with Al/AlO$_x$/Al tunnel
junctions were fabricated on Si substrate buffered by a
sputtered Al$_2$O$_3$ layer 200 nm thick. Shadow
evaporation\cite{Niem-Dol} at three or four angles was used. The
peculiarity of our method was that each electrode (the
base and counter electrodes and the island) was formed individually.
\begin{figure}[h]
\begin{center}
\leavevmode
\includegraphics[width=3in]{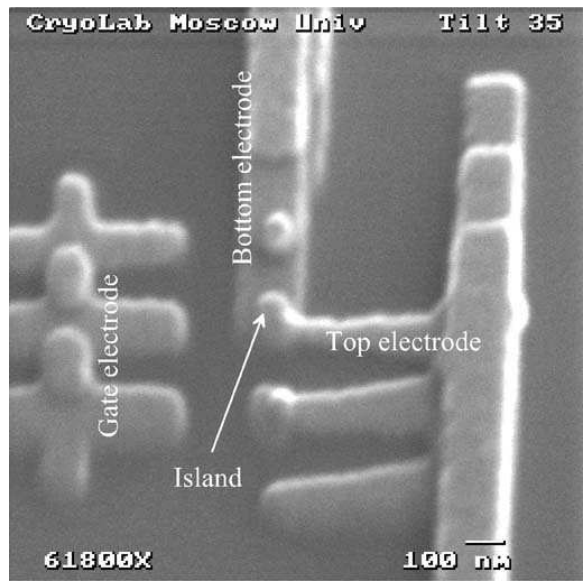}
\caption {SEM image of the typical transistor of stacked geometry.
Triple shadows resulted from three successive depositions of Al
through the same mask. } \label{fig:fig1}\end{center}\end{figure}

We fabricated and studied the
three three types of samples. The transistors of
type I (Fig.~2a-d) were fabricated in three
deposition steps {\it in situ}. After the first and second  depositions
the structure was oxidized to form tunnel barriers of the
transistor junctions. We used a series of masks which allowed a
gradual transformation from the planar transistor structure
(Fig.~2a) to the stack geometry (Fig.~2d) to be realized. For
stack transistors, particular attention was paid to precisely
aligning the island and the base electrode to avoid contact
between the electrodes.
\begin{figure}[h]
\begin{center}
\leavevmode
\includegraphics[width=5in]{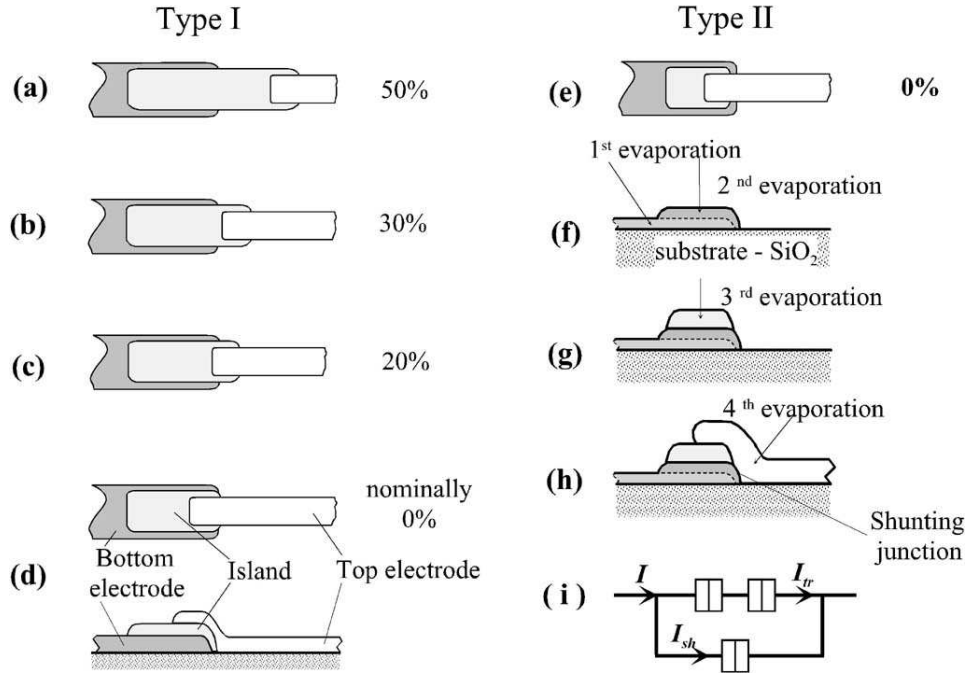}
\caption {Geometry of the series of the type I transistors with
different contact area between the transistor island and the
substrate (a-d). The contact area is expressed as percentage of
the whole island area. The junctions of type I transistors are of
almost identical sizes. Their electric parameters are
$R_{\Sigma}=200-450 $k$\Omega $, $C_{\Sigma}=350-450$~aF,
 $C_1/C_2\approx R_2/R_1\approx 3-5$ and
$C_{\Sigma} \sim 0.8-0.2$~aF. The transistor of
type II (e), the sequence of fabrication steps (f-h) and the
equivalent electric circuit (i) are shown. In this sample, the island (see
also\cite{Denis}) is completely placed on the base electrode. The
device has a small shunting tunnel junction ($R_{sh} \approx 1.9
$M$\Omega $). The electric parameters of the transistor are
$R_{\Sigma }=R_1+R_2 \approx 3.9 $M$\Omega $,
$C_{\Sigma}=C_1+C_2+C_g \approx 270$~aF,
$C_{\Sigma} \approx 0.2$~aF and $C_1/C_2\approx
R_2/R_1\approx
20-30$.}\label{fig:fig2}\end{center}\end{figure}

In contrast to type I, the transistor of type II (Fig.~2e) was
intentionally fabricated to have a shunting tunnel junction
between the outer electrodes. This was the price to be paid
for the complete
and reliable isolation of the island from the substrate. The base
electrode of this device was deposited (Fig.~2f) in two steps,
without oxidation between depositions. The island was formed in
the third (Fig.~2g) deposition step at the same angle and through
the same opening in the mask as in the previous (second)
deposition. This self-aligned method and the fact that the mask
opening shrinks during deposition guarantees that the island is
deposited exactly on top of the base electrode with no contact to
the substrate. As a result, a shunting junction between the base
and counter electrodes was formed after the last, fourth
deposition (Fig.~2h). The presence of the shunting junction has
testified the absence of contact between transistor island
and substrate.

The design of the sample of type III is depicted in Fig.~3. This
transistor also has a tunnel junction shunt and the island
is not in contact with the substrate. The structure consists of
two touching fingers (electrodes) with an island positioned on top
of these fingers. It has been fabricated by shadow evaporation
at three different angles, with an oxidation after the first and
second deposition steps.
\begin{figure}[h]
\begin{center}\leavevmode
\includegraphics[width=5in]{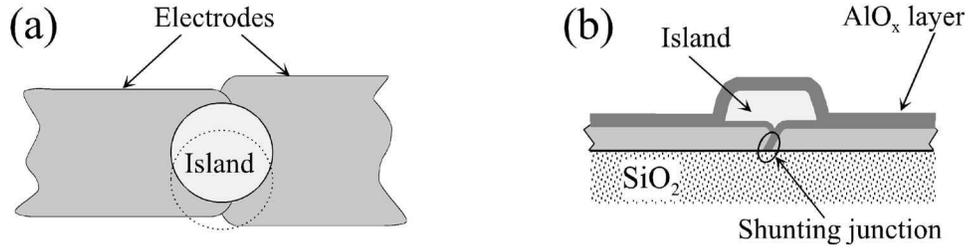}
\caption {Top view (a) and cross-section (b) of the type III
transistor with an island placed onto the outer electrodes, thus
isolated from the substrate. The device has the
small-size shunting tunnel junction ($R_{sh} \approx 1.1\
$M$\Omega $). The electric parameters of the transistor are
$R_{\Sigma }=R_1+R_2 \approx 0.75\ $M$\Omega $ and
$C_{\Sigma}=C_1+C_2+C_g \approx 250\ $aF. A dotted line shows a
possible shift of the transistor island from its nominal position
(solid line), giving rise to small-area contact between
island and substrate.}
\label{fig:fig3}\end{center}\end{figure}

\section{EXPERIMENT}

The electric and noise characteristics of all samples were
measured in a dilution refrigerator at the bath temperature of
$T=25$~mK. The magnetic field $B=1$~T was applied to suppress
superconductivity in the Al films. A voltage bias configuration
was chosen and an output current $I$ measured by a
transimpedance amplifier\cite{Starm1}.

To observe the noise influence of the dielectric substrate on the
electrometer performance, we have measured and compared the
equivalent charge noise figures of the set of our devices. In most
of the cases the magnitude of the noise signal  depended on a
slope $dI/dV_g$ of the modulation curve at a working point (see,
for instance\cite{Starm2}), pointing to the charge nature of
the noise. We measured the low frequency ($f<100$~Hz) noise
spectra and characterized our samples by the magnitude of the
charge fluctuations at $f=10$~Hz. The results are presented in
Figs.~4-6.
\subsection{Devices of type I}

The transistors of type I and type II, both of stacked
geometry, had the lowest noise figure.
\begin{figure}[h]
\begin{center}\leavevmode
\includegraphics[width=3.4in]{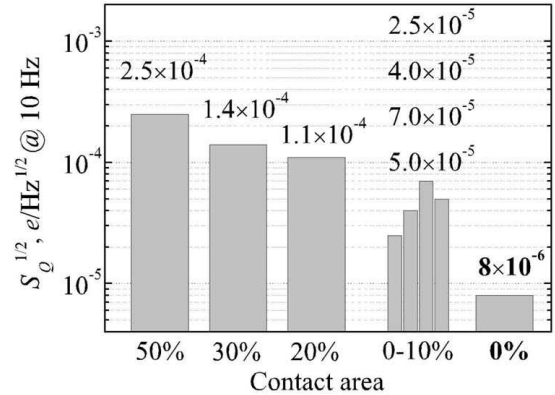}
\caption {Charge noise in the transistors of different geometries
(shown in Fig.~2a-e) measured at small currents ($I\approx 10-
20$~pA) and low frequency ($f=10$~Hz). The transistors with the
contact areas of 50\%, 30\% and 20\% (Fig.~2a-c) and one of the
stacked
transistors (Fig.~2d), with a noise level of $7 \times 10^{-5}
e/\sqrt{\mbox{Hz}}$, were fabricated on the same chip.}
\label{fig:fig4}\end{center}\end{figure}The diagram in Fig.~4
demonstrates the gradual decrease of the charge noise in the set
of transistors (type I) with decreasing the island-substrate
contact area. Since the perfect Coulomb blockade was observed in
all stack transistors of type I investigated (Fig.~2d), i.e. they did
not show any sign of a shunt between the base and counter
electrodes, we cannot exclude the existence of small areas
where the edge of the island is in contact with the substrate.
These areas potentially contribute to the total noise of
transistors, and this could explain the noticeable difference in the
noise levels (see Fig.~4) of these devices.

The lowest charge noise
level among the series of stacked transistors of type I, measured
at $f=10$~Hz, was found to be $2.5\times 10^{-5}e/
\sqrt{\mbox{Hz}}$ or, in energy units, $\sim230\hbar
$ \cite{vkrup2}. Note that this level is still considerably higher
than the fundamental white noise floor whose value estimated\cite{kor2}
for our case is about $3\times 10^{-6}e/\sqrt{\mbox{Hz}}$ or $\sim
4\hbar$~\cite{remark}. On the other hand, the noise figure obtained  is
substantially lower than the best one obtained for an electrometer of
traditional (planar) geometry: $7\times 10^{-5}e/\sqrt{\mbox{Hz}}$
or $\sim 1000 \hbar$ \cite{Visscher}.

\subsection{Devices of type II}

The result obtained for the transistor of type II turned out to be
superior to that obtained for the series of stacked transistors of type I
(see Fig.~4).
\begin{figure}[h]
\begin{center}
\includegraphics[width=5in]{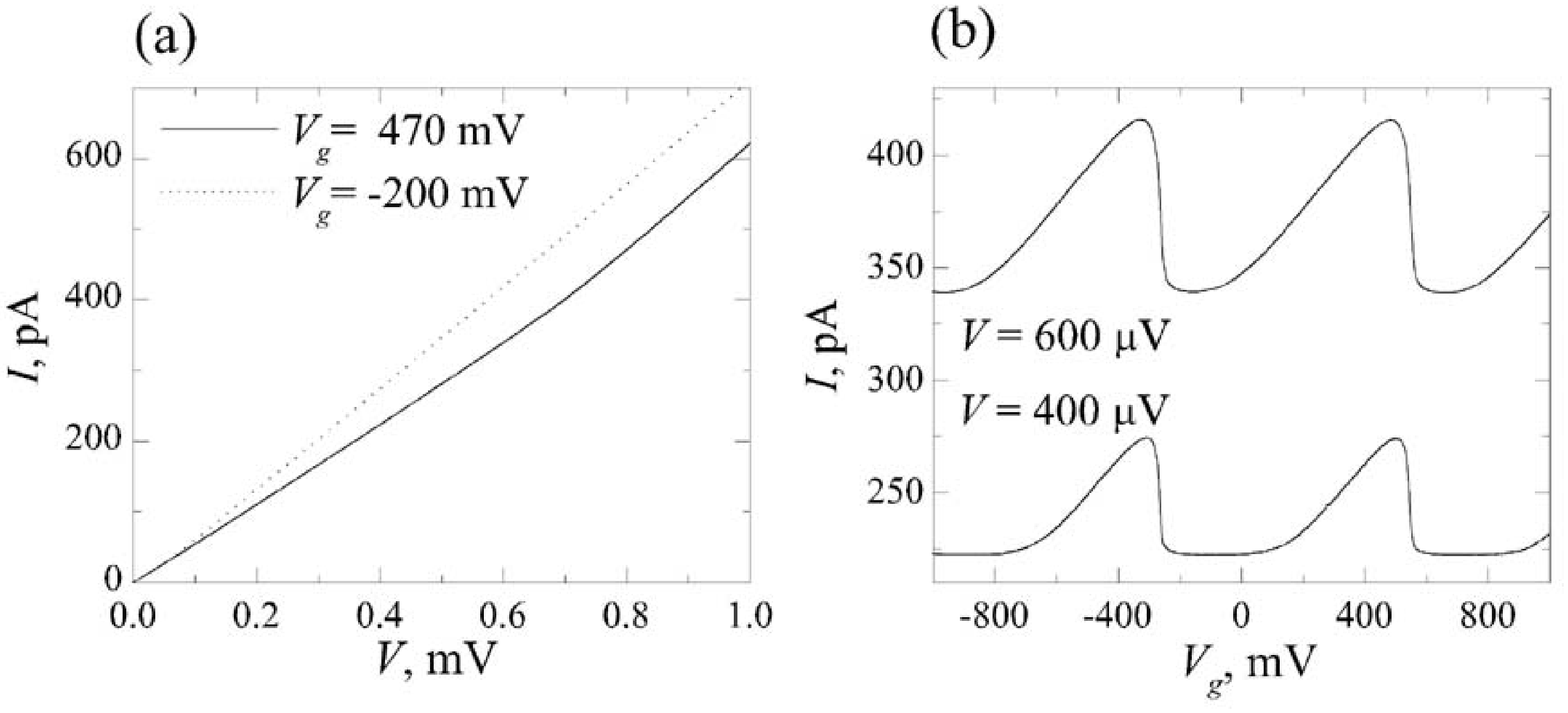}
\caption {Typical $I-V$ (a) and $I-V_g$ (b) curves of the type II
transistor.  $I-V$ curves are presented in the blockade (solid
line) and open (dotted line) states of the transistor. The
presence a non zero conductance between the counter electrodes
leads to a finite slope of the $I-V$ curve in the blockade state
of the transistor and to the shift of $I-V_g$ curves depended on
voltage bias $V$.} \label{fig:fig5}\end{center}\end{figure}
\begin{figure}[h]
\begin{center}\leavevmode
\includegraphics[width=5in]{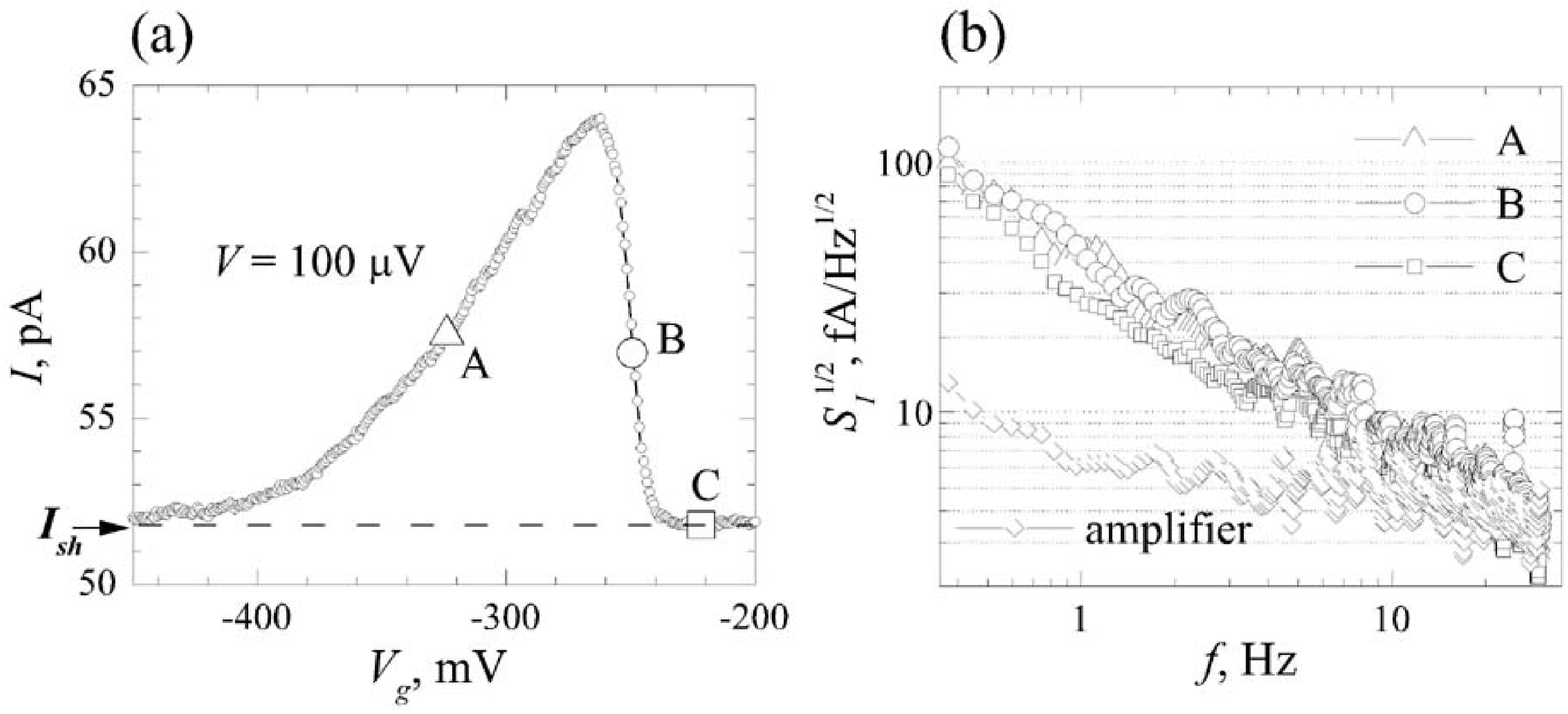}
\caption{$I-V_g$ curve (a) and the output noise spectra (b)
measured in the sample of type II at small current (6~pA) and at
points (A, B, C) with different values of the current-to-charge
ratio $dI/dQ_0$, where $Q_0=C_gV_g$.}
\label{fig:fig6}\end{center}\end{figure} The presence of the
shunting junction strongly modified the $I-V$ curves within the
Coulomb blockade range (see Fig.~5a), but it did not hinder the
gate modulation of the transistor current (Fig.~5b). The total
current ($I=I_{tr}+I_{sh}$) through the structure (see Fig.~2i)
consists of two components: the first ($I_{tr}$) depends
strongly on the gate and bias voltages ($V_g$ and $V$) whereas the
second ($I_{sh}$) depends linearly on bias voltage $V$ only.
This fact resulted in that the $I-V$ curve of the device (Fig.~5a)
was a superposition of the $I-V$ curve of SET transistor and a
linear $I-V$ curve of the single junction. The only modification
of the resulting $I-V_g$ curves (Fig.~5b) was their additional
vertical shift of $I_{sh}=V/R_{sh}$.

At sufficiently low $I$ (corresponding to $I_{tr}=5-10$pA),
the noise level (see Fig.~6) did not show substantial dependence on
the gate voltage, while at large $I$ the magnitude of the output
noise depended, as is usual for the planar devices, on the value of
the transfer function $dI/dV_g$ in a working point indicating
the charge nature of the noise in the latter case. The fact that in
the former case the difference between the noise spectra measured
at working points A, B and C in Fig.~6b was insignificant,
indicated that the charge noise component in the total
transistor noise was not dominant. The evaluation of the noise
contribution of the amplifier
(${\delta}I_{amp}=5-6$~fA$/\sqrt{\mbox{Hz}}$) to the total noise
signal (${\delta}I=8\pm 2$fA$/\sqrt{\mbox{Hz}}$) led to a very low
value of the electrometer noise related to the input charge:
${\delta}Q_x=(8\pm 2)\times 10^{-6}e/\sqrt{\mbox{Hz}}$ or, in
terms of energy sensitivity, $\sim30\hbar $ at 10 Hz~\cite{remark}. This
value
is even below the level ($1.2\times 10^{-5}e/\sqrt{\mbox{Hz}}$ or
$41\hbar$) recently attained by the so-called rf-SET
transistor\cite{Schoel} at much higher ($f = 1.1$~MHz) frequency
where the offset charge noise should be obviously roll-off.
\begin{figure}
\begin{center}\leavevmode
\includegraphics[width=3.4in]{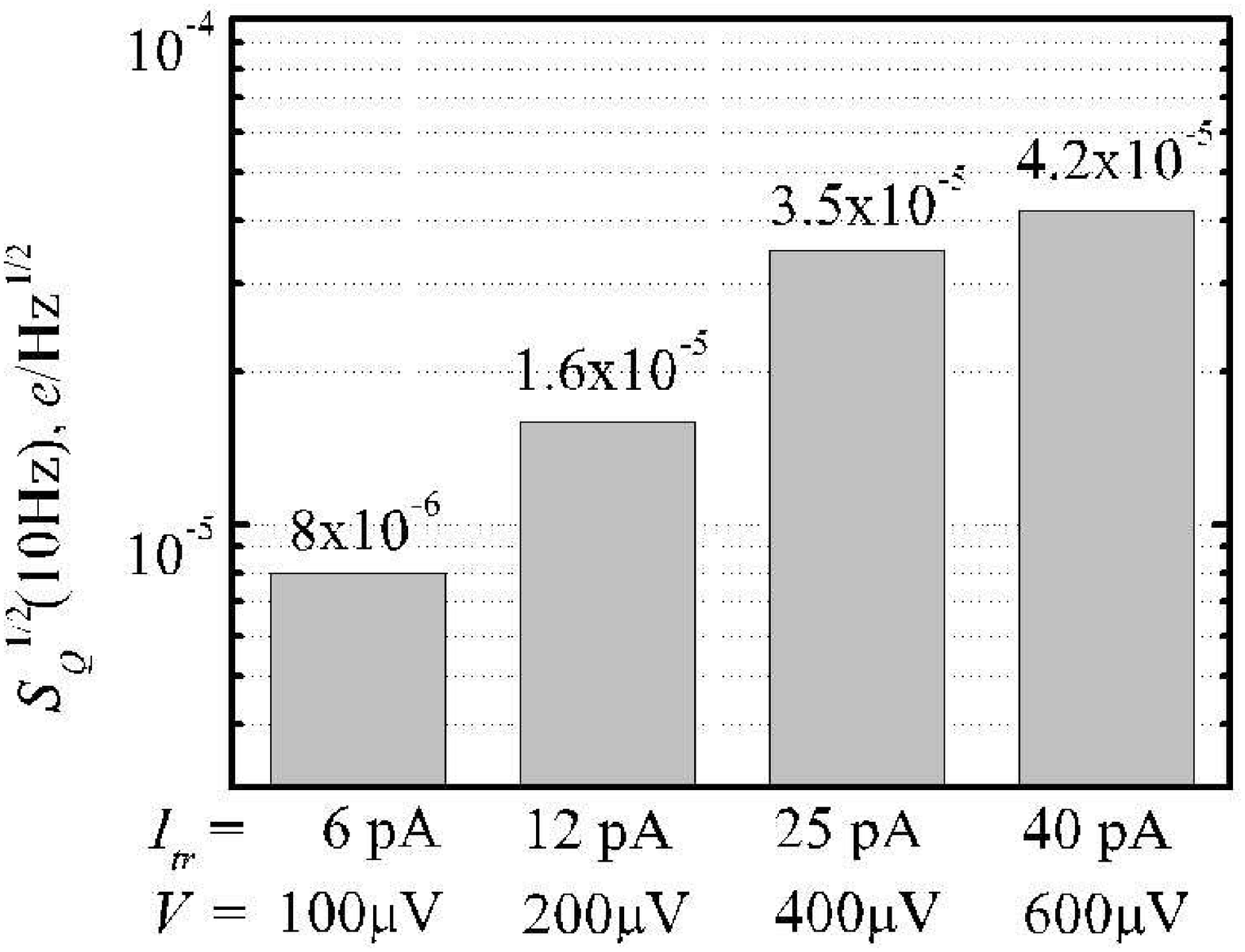}
\caption { The charge noise of the type II transistor obtained for
different biases $V$ at frequency $f=10$~Hz.}
\label{fig:fig7}\end{center}\end{figure} Moreover, the lowest
noise figure measured in a "shunted" stacked transistor at 10 Hz is
close to its fundamental white noise floor\cite {kor2} whose level
for this sample was estimated to be ${\delta}Q_x=(2\div3)\times
10^{-6}e/\sqrt{\mbox{Hz}}$ or $3\hbar$. Since the measured noise
magnitude still shows a trend to decrease with frequency, we
believe that the electrometer sensitivity can approach its
fundamental noise limit at frequencies of some ten Hz. The extremely
small signal corresponding to the white noise floor could be
measured with the aid of a less noisy (presumably cold) amplifier.

\subsection{Devices of type III}

The charge noise level of the transistor of type III (Fig.~3) was
also found to be relatively low: ${\delta}Q_x=7\times
10^{-5}e/\sqrt{\mbox{Hz}}$ at $f=10$Hz, although it is
considerably higher than that of the type II transistor. We
believe that this is because of incomplete screening of the island
from noise sources of the substrate. We assume that  the island
was placed  so that its small part contacted the substrate (as
shown in Fig.~3 by dotted line) because of imperfect alignment.
Nevertheless, this design has obvious advantages: first, it is
easier to fabricate than that of devices of types I and II and
secondly, what is even more important, it gives the opportunity to
reduce further the dimensions of the tunnel junctions and, hence,
to increase an operation temperature of the SET electrometer.

\section{DISCUSSION}

Our experiments with metallic SET transistors of different design
clearly show that the dominant contribution to the
background charge noise is associated with a "noisy" substrate.

The stacked samples in which the island-substrate contact area was
minimized (devices of types I, II and III) exhibit pretty low
noise at low transport current. In some cases (transistor of type
II) this residual low-frequency noise was insensitive to the gate
voltage and this behavior might be associated with fluctuations of
junction conductance\cite{vkrup2}. At high transport currents and
bias voltages, the
usual gate dependence of noise in these samples is restored and
the noise level increases (see Fig.~7). Such behavior can be
explained by activation of the charge traps inside the dielectric
barriers and the natural Al oxide covering the whole sample
surface. In particular, the perimeter area of the island seems to
be most sensitive to random recharging of traps located nearby,
because a charge induced on a metallic surface of small radius
of curvature strongly depends on a distance to the source-charge.
As regards the tunnel barriers, their charge noise behavior (if
any) at low currents $I_{tr}$ is remarkable. Even for very small
$I_{tr}$, the electric field inside the barriers oscillates with
the SET rate $I_{tr}/e$ and amplitude of $A = e/(dC_\Sigma)$,
where $d$ is the barrier thickness\cite {Zor}. However, this
rather strong field does not produce an appreciable random
switching of the barrier traps. On the other hand, a strong
alternative field can possibly re-charge the potential traps with
the rate of SET oscillations.

In conclusion, we have proposed two possible ways (devices of
types II and III) which allow the noise
characteristics of SET devices to be drastically improved.
In particular, the obtained charge
noise level of the SET electrometer, with an island isolated from a
substrate, ${\delta}Q_x=(8\pm2)\times 10^{-6}e/\sqrt{\mbox{Hz}}$ is
the lowest one ever reported. The noise measurements of the
device which island are not in contact with the substrate give
justified hope that SET devices of traditional planar design with
Al/AlO$_x$/Al tunnel barriers can potentially have much lower
offset charge noise provided that an appropriate (noiseless) material
for the substrate can be found.

\section{ACKNOWLEDGMENTS}

This work has been supported in part by the German BMBF, the EU Project
CHARGE, the Russian Scientific Program "Physics of Solid State
Nanostructures" and the Russian Foundation for Basic Research.


\begin{thebibliography}{99}
\bibitem{remark} We characterize the sensitivity of our devices in terms
of equivalent charge noise referred to the transistor island.
\end{thebibliography}
\end{document}